\documentclass[aps,prd,showpacs,twocolumn,superscriptaddress,sort&compress]{revtex4-1}

\usepackage{graphicx}
\usepackage{amsmath}

\usepackage{microtype}

\begin{document}

\title{Masses of $J^{PC}=1^{-+}$ exotic quarkonia in a Bethe-Salpeter-equation approach}


\author{T. Hilger}
\affiliation{Institute of Physics, University of Graz, NAWI Graz, A-8010 Graz, Austria}

\author{M. G\'{o}mez-Rocha}
\affiliation{Institute of High Energy Physics, Austrian Academy of Sciences, A-1050 Vienna, Austria}
\affiliation{Institute of Physics, University of Graz, NAWI Graz, A-8010 Graz, Austria}

\author{A. Krassnigg}
\email[]{andreas.krassnigg@uni-graz.at}
\affiliation{Institute of Physics, University of Graz, NAWI Graz, A-8010 Graz, Austria}

\date{\today}

\begin{abstract}
We investigate the properties of mesons with the exotic $J^{PC}=1^{-+}$ quantum numbers.
Starting out from the light-quark domain, where the $\pi_1$ states are used as references,
we predict the masses of analogous quarkonia for $c\bar{c}$ and $b\bar{b}$
configurations. We employ a covariant Dyson-Schwinger-Bethe-Salpeter-equation approach with a
rainbow-ladder truncated model of quantum chromodynamics.
\end{abstract}


\maketitle

\section{Introduction\label{sec:intro}}


After its origin several decades ago, hadron spectroscopy has seen increased interest
in recent years. In particular, states not predicted or readily explained by the comprehensive
quark-model calculations of the past, e.\,g., \cite{Godfrey:1985xj} have sparked
both theoretical investigations as well as experimental searches at modern and future
hadron-physics facilities; foremost, in the present manuscript we focus on
mesons with exotic quantum numbers, which have been reviewed in great detail, e.\,g., very recently
\cite{Meyer:2015eta}.

The approach we chose is the coupled Dyson-Schwinger--Bethe-Salpeter-equation (DSBSE) framework for
the investigation of quantum chromodynamics (QCD),
a well-established covariant variant \cite{Roberts:2007jh,Fischer:2006ub,Alkofer:2000wg,Sanchis-Alepuz:2015tha} from a 
set of continuum quantum-field-theoretical methods 
\cite{Lucha:2014xla,Wambach:2014vta,Pawlowski:2005xe,Brodsky:2014yha} complementary to the
well-known lattice-regularized QCD \cite{Dudek:2007wv,Liu:2012ze,Thomas:2014dpa}.

Concretely, our present work is built on top of recent studies of 
heavy quarkonia \cite{Blank:2011ha,Hilger:2014nma}, to which we add
an immediate focus on the exotic meson quantum numbers $J^{PC}=1^{-+}$, thus approaching
the quarkonium spectrum from a slightly different perspective. In particular, the present study
reaches out to the light-quark domain for anchoring in addition to the conventional quarkonium 
spectra in order to make use of experimental evidence presently at hand for the $\pi_1$ states.
Our setup employs the basic but symmetry-preserving rainbow-ladder (RL) truncation
to study mesons by solving the quark Dyson-Schwinger equation (DSE) coupled
to the meson $q\bar{q}$ Bethe-Salpeter equation (BSE). 

%

\section{Setup and Model\label{sec:setup}}

While we perform an RL-truncated study of Landau-gauge QCD in Euclidean space using the homogeneous BSE, 
we refer the reader to alternate routes of studies such as ours via a brief list of exemplary references
to Coulomb-gauge QCD \cite{Alkofer:2005ug,Rocha:2009xq,Cotanch:2010bq,Popovici:2011yz}, 
the BSE in Minkowski space \cite{Carbonell:2014dwa,Frederico:2013vga,Sauli:2012xj}, 
an analogous but more general inhomogeneous vertex BSE \cite{Maris:2000ig,Bhagwat:2007rj,Blank:2010sn},
and studies beyond RL truncation \cite{Williams:2014iea,Gomez-Rocha:2014vsa,Sanchis-Alepuz:2014wea,Chang:2009zb,Heupel:2014ina},
from which the relevant literature can be grasped. 

The strength of the setup chosen herein is its combination of computational and 
conceptional feasibility \cite{Krassnigg:2008gd,Blank:2010bp}
with the important features inherent to a systematic setup of a nonperturbative quantum-field-theoretical method. 
In particular, in QCD the implementation of chiral symmetry and its dynamical breaking as well as a realization
of confinement and meson properties expected in the heavy-quark limit are paramount. Our study meets these
requirements via properties demonstrated of RL truncated DSBSE studies with dressed quarks before, namely 
satisfaction of the axial-vector Ward-Takahashi identity \cite{Munczek:1994zz,Maris:1997hd,Holl:2004fr}, 
quark-confinement \cite{Alkofer:2003jj,Alkofer:2008tt}, 
and adequate behavior towards the heavy-quark limit \cite{Krassnigg:2009zh,Maris:2005tt,Nguyen:2010yh}.

We use the homogeneous $q\bar{q}$ BSE in RL truncation which reads 
\begin{eqnarray}\nonumber
\Gamma(p;P)&=&-\frac{4}{3}\!\!\int^\Lambda_q\!\!\!\!\mathcal{G}((p-q)^2)\; D_{\mu\nu}^\mathrm{f}(p-q) \;
\gamma_\mu \; \chi(q;P)\;\gamma_\nu \\ \label{eq:bse}
\chi(q;P)&=&S(q_+) \Gamma(q;P) S(q_-)\,,
\end{eqnarray}
where $q$ and $P$ are the quark-antiquark relative and total momenta, respectively, and 
the (anti)quark momenta  are chosen as $q_{\pm} = q\pm P/2$.
The renormalized dressed quark propagator $S(p)$ is obtained from the corresponding DSE 
\begin{eqnarray}\nonumber
S(p)^{-1}  &=&  (Z_2 i\gamma\cdot p + Z_4 m_q(\mu))+  \Sigma(p)\,,\\\label{eq:dse}
\Sigma(p)&=& \frac{4}{3}\!\!\int^\Lambda_q\!\!\!\! \mathcal{G}((p-q)^2) \; D_{\mu\nu}^\mathrm{f}(p-q)
\;\gamma_\mu \;S(q)\; \gamma_\nu \,.
\end{eqnarray}
$\Sigma$ is the quark self-energy, $m_q$ is the current-quark mass chosen at a renormalization scale $\mu$,
$D_{\mu\nu}^\mathrm{f}$ represents the free gluon propagator and $\gamma_\nu$ is
the bare quark-gluon vertex's Dirac structure.  Dirac and flavor indices are omitted for brevity.
$\int^\Lambda_q=\int^\Lambda d^4q/(2\pi)^4$ denotes a
translationally invariant regularization of the integral, with the regularization scale
$\Lambda$ \cite{Maris:1997tm}.

In a given truncation the effective interaction $\mathcal{G}$ needs to be specified according to
the aims of the study. We use the well-investigated and phenomenologically successful parameterization 
of Ref.~\cite{Maris:1999nt}, which reads
\begin{equation}
\label{eq:interaction} 
\frac{{\cal G}(s)}{s} =
\frac{4\pi^2 D}{\omega^6} s\;\mathrm{e}^{-s/\omega^2}
+\frac{4\pi\;\gamma_\mathrm{m} \pi\;\mathcal{F}(s) }{1/2 \ln
[\tau\!+\!(1\!+\!s/\Lambda_\mathrm{QCD}^2)^2]}.
\end{equation} 
The parameter $\omega$ corresponds to an effective
inverse range of the interaction, while $D$ acts like an overall strength of the
first term; they determine the intermediate-momentum part of the 
interaction, while the second term is relevant for large momenta and produces the
correct one-loop perturbative QCD limit. We note that
${\cal F}(s)= [1 - \exp(-s/[4 m_\mathrm{t}^2])]/s$ where $m_\mathrm{t}=0.5$~GeV,
$\tau={\rm e}^2-1$, $N_\mathrm{f}=4$, $\Lambda_\mathrm{QCD}^{N_\mathrm{f}=4}=
0.234\,{\rm GeV}$, and $\gamma_\mathrm{m}=12/(33-2N_\mathrm{f})$, which is 
left unchanged from Ref.~\cite{Maris:1999nt}.

The parameters considered independent and free in our study are $\omega$ and $D$ (in addition to the 
current-quark masses), which we already used previously to achieve a surprisingly good phenomenological
description of meson spectra \cite{Popovici:2014pha,Hilger:2014nma} in the heavy-quark domain. 
In the present study we refit the pair $\omega$ and $D$ using the same quark masses as in 
\cite{Popovici:2014pha,Hilger:2014nma} in order to specifically target the $J^{PC}=1^{-+}$ channel.
Further details are specified below.

\section{Exotic Quantum Numbers\label{sec:exotics}}

Exotic quantum numbers have been addressed within the DSBSE approach several times before 
\cite{Krassnigg:2009zh,Burden:1996nh,Burden:2002ps,Watson:2004kd,Qin:2011xq,Fischer:2014cfa,Rojas:2014aka,Jarecke:2002xd,Hilger:2014nma},
but have been in the actual focus of the work only in \cite{Burden:2002ps,Qin:2011xq}.

%
In a short critique of Ref.~\cite{Qin:2011xq} we would like to mention here in particular the claim that
$\omega$ and $D$ are not independent so that ``one can expect computed observables to be
practically insensitive to $\omega$  on the domain $\omega \in$ [0.4, 0.6] GeV.'' In our opinion,
this is an unnecessary and over-restrictive assumption which actually prevents part of the
possibilities a complete set of results may have. We note that there is a difference between having
a one-parameter model at hand like in older studies to investigate effects the long-range behavior of the
strong interaction might have on states of various kinds \cite{Holl:2004un,Holl:2005vu,Krassnigg:2009zh,Krassnigg:2010mh} 
on one hand, or attempting a complete
study of the possibilities within the parameter space of a given model interaction on the other hand. While in \cite{Qin:2011xq} 
the former is present and, in fact, a different value for the product of $\omega D$ is used for excited and exotic states to
attempt a description along the lines of beyond-RL premises, the latter is not, and so any conclusions drawn there are 
necessarily incomplete. For example, general statements about excited and exotic states are made in \cite{Qin:2011xq} 
based on light-quark results only, while heavy mesons are not considered, nor are those with axial-vector and tensor 
quantum numbers. In our present work, on the contrary, we show that axial-vector states may be the key to a proper 
understanding of, e.\,g., the exotic vector states in the light-meson spectrum.

In another short critique, namely of Ref.~\cite{Burden:2002ps}, we remark that, after an in-depth discussion
of the DSBSE setup and it's capabilities to describe states with exotic quantum numbers, the authors employ
a strategy similar to ours. In particular they focus on states with $J=1$, together with the obligatory 
pseudoscalar ground state, and obtain reasonable agreement with experimental data for vector, axial vector, and
exotic vector ground states. Since the choice of a separable kernel in the BSE limits the resulting equations
and bound-state amplitudes, this set of results cannot easily be generalized or extended. For example, only two 
solutions are supported in the $1^{-+}$ channel which also appear rather close together in mass, and an immediate 
comparison to the RL setup employed herein is difficult. At the time, possible interpretations as to why the results in 
\cite{Burden:2002ps} compared more favorably to the experimental situation than RL studies were mainly that
there was lesser, and thus more appropriate, strength in the separable as compared to the RL kernel for light quarks.

With these two DSBSE studies of states with exotic quantum numbers in mind, we now very briefly sketch
a few relevant issues before outlining our strategy towards meaningful statements in our approach.

In the DSBSE approach, states with exotic quantum numbers appear naturally in the setup of the quark-antiquark BSE. 
Owing to the ubiquity of the scalar product of the quark-antiquark relative and total momenta, $q\cdot P$, and its 
negative $C$-parity, the appearance of such states is a direct result of the dynamics and interaction considered.
While in lattice QCD or constituent-quark models some explicit inclusion of gluonic degrees of freedom leads to the
exotic value of $C$, a similar mechanism includes gluonic dynamics here \emph{implicitly}; for more details and
a thorough discussion, see \cite{Burden:2002ps}. For the present purpose and possible interpretations of our results,
it is mainly important to remark here that our setup does not actually imply a non-hybrid interpretation.

Regarding the experimental situation of states seen in the $1^{-+}$ channel, we are aware of an ongoing
discussion about the nature and properties of the $\pi_1$ states as they are listed in \cite{Olive:2014rpp}.
However, improvements are expected due to the wide efforts undertaken both experimentally and theoretically to 
make the overall quality and understanding of the hadron spectrum even better \cite{Battaglieri:2014gca}.
For the purposes of our argument, we assume enough credibility of experimental information on the $\pi_1(1400)$ and 
$\pi_1(1600)$ states to warrant our hypothesis, which is certainly the case. The $\pi_1(2015)$ is neither
needed for our fit setup, nor matched by any of our resulting states (our results are too high to match it) 
and so is omitted from our discussion.

\begin{figure*}
\centering
 \includegraphics[width=0.75\textwidth]{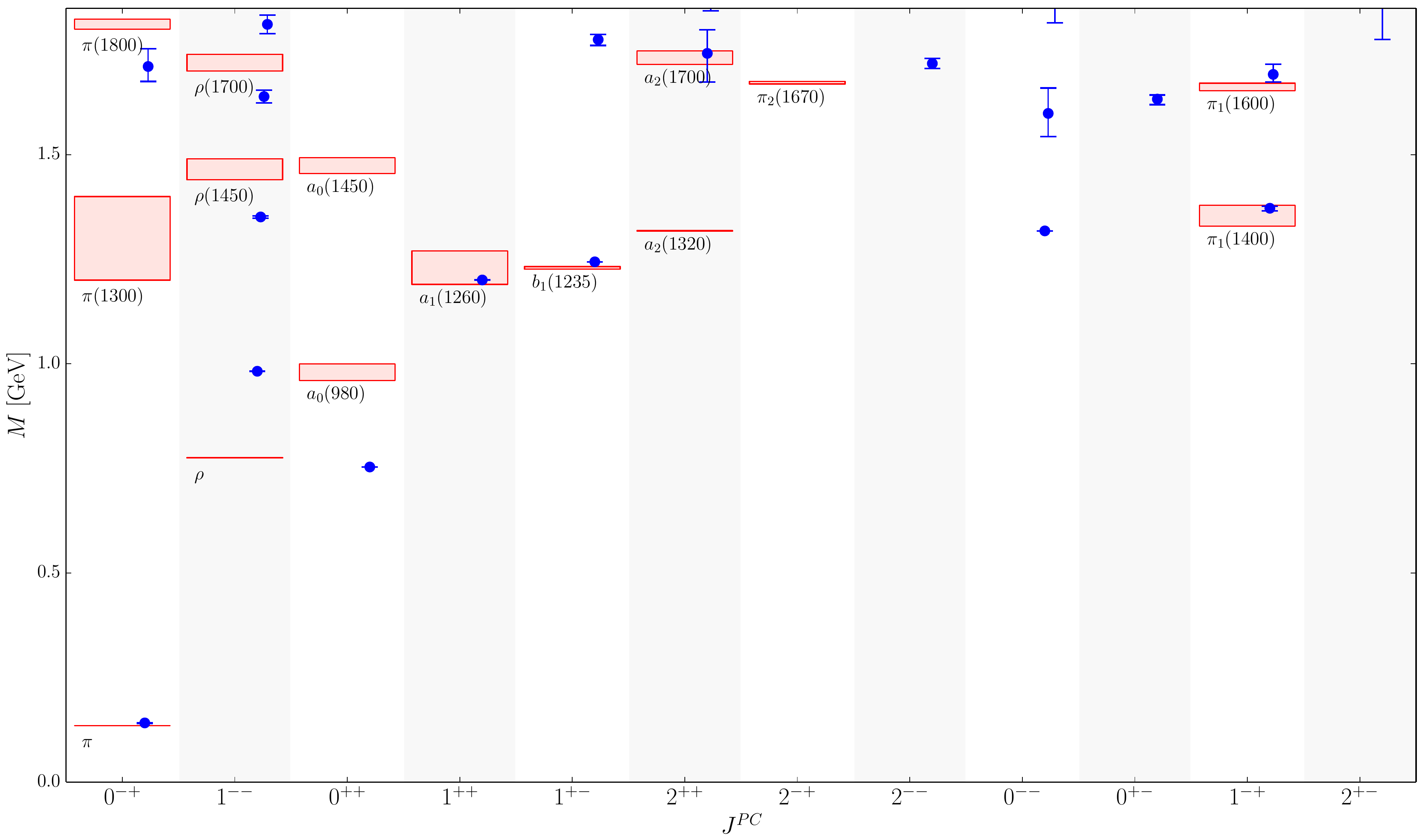}
\caption{Light isovector quarkonium spectra of mesons with $J\le 2$ including exotic quantum
numbers, fitted to exotic-vector-related splittings. Calculated values (blue circles) are compared to experimental data (red boxes) from 
\cite{Olive:2014rpp}.}\label{fig:isov-spectrum-exotic}      
\end{figure*}

\section{Strategy\label{sec:strategy}}

Naturally, a study using a truncated set of equations like ours has to be taken with a grain of
salt. As a result, we employ the following strategy in pursuit of meaningful predictions for $1^{-+}$ meson
masses in the quarkonium spectrum: While an overall successful description of states with $J\le 2$
in RL truncation has been proven feasible in the heavy-quark domain \cite{Hilger:2014nma}, there
is no reason \emph{a priori} to assume similar success in the light-quark sector. On the contrary,
a comprehensively successful phenomenological description of the same quality, even in the 
isovector case, would be a surprise. Nonetheless, subsets of observables can be targeted by
informed and suitably eclectic studies, which is our tactic herein as well. 

\begin{figure*}
\centering
 \includegraphics[width=0.75\textwidth]{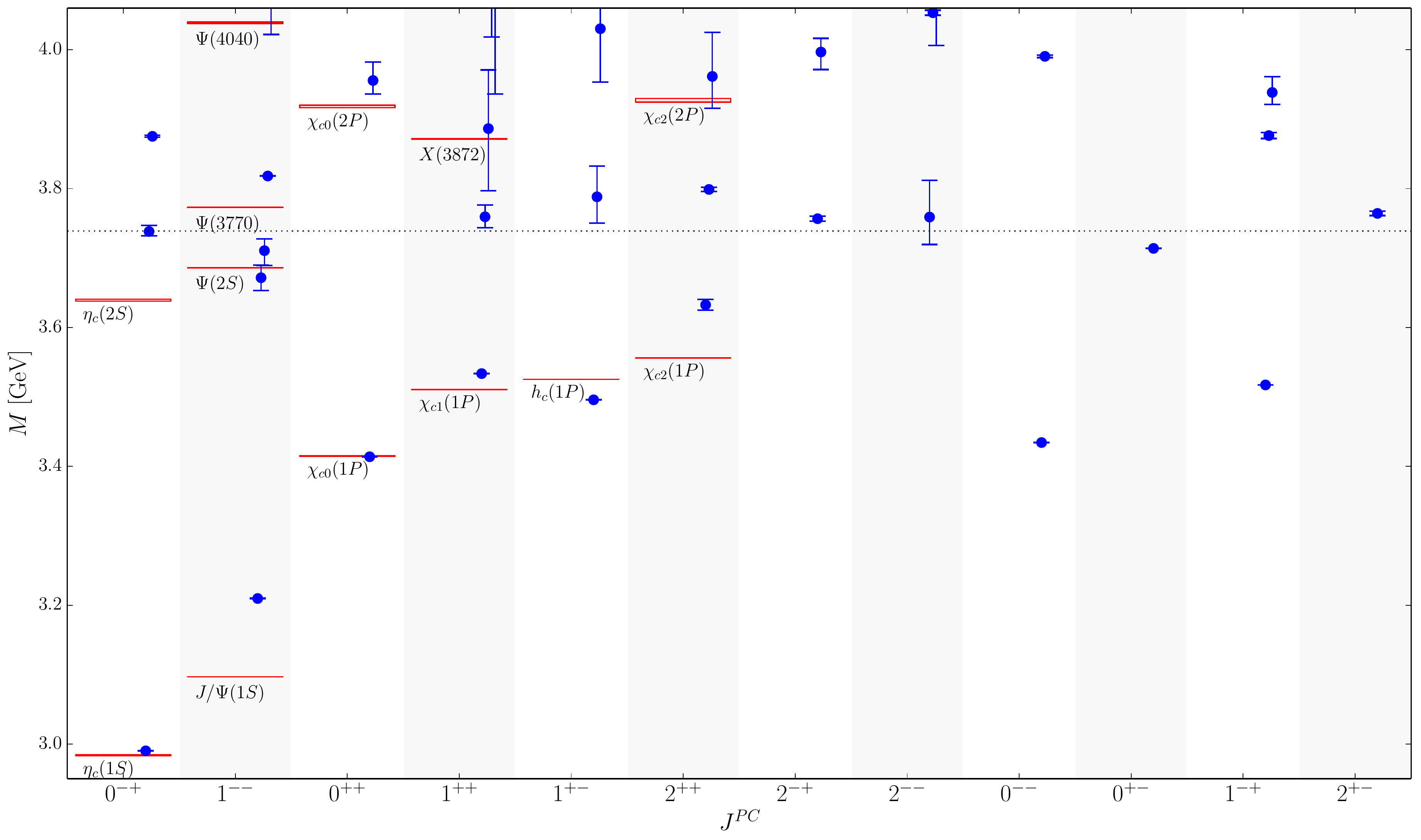}
\caption{Charmonium spectra, fitted to axial-vector-related splittings. Calculated values (blue circles) are 
compared to experimental data (red boxes) from 
\cite{Brambilla:2010cs,Olive:2014rpp,Bhardwaj:2013rmw,Ablikim:2015baa}. The horizontal dashed line marks the lowest open-flavor threshold.}\label{fig:isov-spectrum-cc}      
\end{figure*}

In particular, as already indicated above, we make use of the experimental information available on the
two lowest-lying exotic $\pi_1$ states, the $\pi_1(1400)$ and $\pi_1(1600)$, to test and constrain those meson 
mass splittings that may be appropriate, i.\,e., well-correlated to identify parameter sets potentially successful
in the description of the $1^{-+}$ channel also for the other quark masses. To find these, we use
comparisons of calculated and experimental mass splittings and of sets of these on a grid
$\omega\in \{0.4,0.5,0.6,0.7,0.8\}$ GeV 
$\times$ $D \in \{0.9,1.1,1.2,1.3,1.4,1.5,1.7\}$ GeV${}^2$ in the isovector case. 
The pion mass is always fixed to its experimental value, which sets the current-quark mass
in Eq.~(\ref{eq:dse}) to $m_{u/d}=0.003$ GeV (given at a renormalization point $\mu=19$ GeV). 
We calculated meson masses for all $J^{PC}$ quantum 
number sets for $J\le 2$ on the grid and investigated the corresponding $\chi^2$ for a number
of comparisons. A more detailed account of this strategy as well as an illustration is 
presented in \cite{Popovici:2014pha}.

To confirm that our region of interest is indeed well-captured by the grid defined above, 
we investigated the $\pi_1$ states and found that a combination of the radial splitting 
between the ground and first excited $1^{-+}$ states, together with the two splittings of each 
$1^{-+}$ state to the ground-state pseudoscalar meson (pion) indeed provides a region of low $\chi^2$ 
in the interior of our grid, which is the necessary prerequisite. We note
that the best fit according to this configuration is given by $\omega=0.7$ GeV and $D=1.4$ GeV${}^2$, 
for which the spectrum is plotted in Fig.~\ref{fig:isov-spectrum-exotic}.
In all figures, our results are represented by blue circles, while experimental data are displayed as wide red boxes, 
where the height of the box indicates the size of the experimental error on the mass. Our error bars, where relevant, come from 
extrapolated results in situations where propagator singularities prohibit a direct calculation; details on the source of
this problem and our extrapolation strategy can be found in the appendices of 
\cite{Krassnigg:2010mh,Blank:2011ha,Dorkin:2013rsa,Hilger:2014nma,Dorkin:2014lxa}.

The next step is to identify a set of non-exotic splittings that provide similar results for our fitting-attempts, 
i.\,e., a reasonable correlation to the exotic case. In search of such a set we have found that
axial-vector states seem to be good indicators. This is not entirely surprising; for example, as mentioned above, 
Ref.~\cite{Burden:2002ps} focused on states from the $J=1$ family and obtained a good overall description,
including exotic states, using a separable kernel for the BSE, while both axialvector and exotic $J=1$ 
states had been difficult to describe otherwise. Another indication for the importance and connection
to axialvector states is the hadronic decay of a $\pi_1$ into axialvector mesons and pions 
\cite{Meyer:2015eta,Burns:2006wz,McNeile:2006bz}, a feature which requires more work on our part and
will be investigated in the future; still, it is important to note already here that a simultaneous good description 
of axialvector, exotic vector, and pion states is an excellent basis for a reliable subsequent calculation
of their hadronic transition amplitude.

With this in mind, we studied several sets of splittings and found that the combination of the ground-state
pseudoscalar to each of the axial-vector ground states together with the intra-axialvector splitting
provide best guidance. The result is $\omega=0.7$ GeV and $D=1.5$ GeV${}^2$. As indicated by the proximity of the 
best-fit parameter sets, the results are indeed very close to each other and we therefore use this configuration
to attempt predictions of $1^{-+}$ states for other quark masses.

\section{Results and discussion\label{sec:results}}

First, we proceed to charmonium, where the same fitting procedure yields 
$\omega=0.6$ GeV and $D=0.9$ GeV${}^2$, for which the spectrum is plotted
in Fig.~\ref{fig:isov-spectrum-cc}. In addition to our results (blue circles) and experimental data (red boxes),
the lowest open-flavor threshold is plotted as a horizontal dashed line. With the pseudoscalar and axial vector
ground states well-met by the fit, we observe further adequate description of the scalar ground state. 
On the other hand, our calculation misses the
vector and (pseudo-)tensor ground states, which is simply a consequence of the different anchoring and targets of the fit, 
as compared to our results in \cite{Hilger:2014nma}. 

On a similar note, one observes that the level ordering of the ground states
in the axialvector $1^{++}$ and $1^{+-}$, while correct for the light-quark
case, is reversed in charmonium. This is a result of a rather straight-forward
dependence of this particular splitting on the model parameter $D$: small values
of $D$ result in the correct level ordering, while increasing $D$ at some point 
flips the level ordering and increases the splitting in the wrong direction. 
In our study, the requirement to reproduce the intra-axial-vector ground-state
splitting is combined with a second requirement, which together yield a good description of the 
exotic light meson spectrum, namely to reproduce the pseudoscalar-axialvector 
splittings. The value of $D$ found via the fit to this combination thus 
determines the axial-vector level ordering and splitting and can miss it by a 
small amount. Apparently, the charmonium result for $D=0.9$ GeV${}^2$ is beyond
the flipping value with the splitting still small, albeit reversed, which is 
acceptable for our purposes.

The lowest charmonium state in the $1^{-+}$ channel has a mass of about
$3.52$ GeV, the next comes out at about $3.88$ GeV.

\begin{figure*}
\centering
 \includegraphics[width=0.75\textwidth]{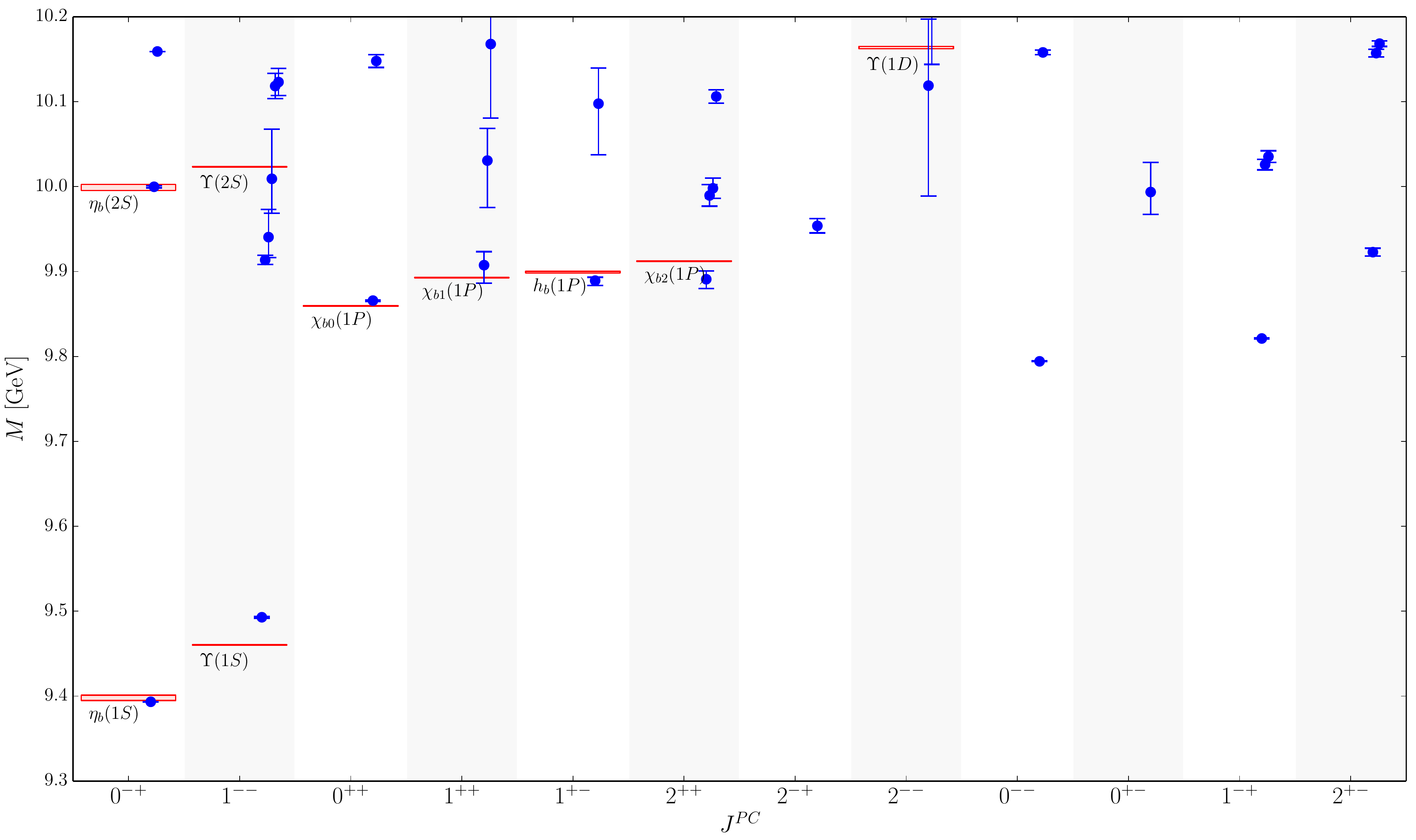}
\caption{Bottomonium spectra, fitted to axial-vector-related splittings. Calculated values (blue circles) 
are compared to experimental data (red boxes) from \cite{Brambilla:2010cs,Olive:2014rpp}. The lowest open-flavor 
threshold is above all masses plotted in this figure.}\label{fig:isov-spectrum-bb}      
\end{figure*}

The situation is similar in the bottomonium case, where our fit arrives at $\omega=0.7$ GeV and $D=0.8$ GeV${}^2$, 
for which the spectrum is plotted in Fig.~\ref{fig:isov-spectrum-bb}. While in this case our description of all
measured ground states is rather accurate, our results for the excited states in many channels seem to be too
low in general. Our resulting $1^{-+}$ ground state appears at $9.82$ GeV and the next at $10.02$ GeV.

While we set out from the isovector $\pi_1$ states, we now return to the light-quark domain for a brief comment on
the isoscalar case. In our RL study, the BSE kernel does not account for flavor-mixing. While one can always mix the
RL-BSE results afterwards \cite{Holl:2004un}, we refrain from doing so here, since there is no straight-forward and
consistent way to implement such a mechanism for our present purposes. In a world with ideal $SU(3)$ flavor mixing
and assumed isospin-symmetry, an RL study provides 
the same results in the light-meson isovector and isoscalar channels. As a consequence, in such a world there should 
also be isoscalar versions of the $\pi_1$ states, based on our simple setup. In addition, at a slightly higher mass,
there would be analogous states with $\bar{s}s$ quark content.

As far as the interpretation of all of these new states is concerned, they may be hybrids
or mix with other types of states, much like in other approaches to this feature of QCD. If one makes use of a 
hybrid interpretation of our results, the immediate question arises, how one would find the partners in possible
corresponding hybrid supermultiplets. However, a study with the ability to identify hybrid and conventional 
states among our results is beyond the scope of our present investigation. On another note, 
we would like to emphasize that our results for the $1^{-+}$ channel, if compared to the results for hybrids 
presently available in the literature \cite{Meyer:2015eta}, are at the lower end of the apparent mass range:
the predicted hybrid $1^{-+}$ masses from lattice QCD, e.\,g., are substatially higher \cite{Liu:2012ze}, whereas
our results are in good agreement with, e.\,g., those from QCD sum rules \cite{Chen:2013zia}.

While we have not aimed to describe states with the exotic quantum numbers $J^{PC}=0^{--}$, $0^{+-}$, or $2^{+-}$,
we do obtain results in these channels and we have included our results for these states in our figures for 
the sake of completeness.
Considering the strategy in our setup specifically anchored to the $J=1$ channels, we have no immediate reason
to expect that our results present reasonable predictions. It is well conceivable that for the various exotic
sets of quantum numbers there are different mechanisms at work and different parts of the interaction responsible
for a good description. Naturally, one can turn to other approaches 
for guidance, which we may explore elsewhere. At present, our focus remains on a straight-forward study
anchored to experimental data.

A final word of discussion concerns the fact that the states in our setup are bound states and not resonances,
since a decay mechanism is not incorporated in the BSE kernel in RL truncation. In particular above the relevant
decay thresholds, this feature must be met with caution. In a first step, a semi-perturbative way to calculate
hadronic decay widths is available and has been tested successfully in the past, see \cite{Mader:2011zf} 
and references therein. Since, in particular, the $\pi_1$ states are resonances, it remains to be seen
whether such a treatment, using our current results, provides also hadronic widths of the correct size.
Furthermore, considerable effects may also be expected from final-state interactions \cite{Bass:2001zs}.

\section{Conclusions and outlook\label{sec:conclusions}}

We presented a covariant study of the masses of prospective $J^{PC}=1^{-+}$ exotic mesons in both
heavy and light quarkonia based on the DSBSE approach to QCD. Starting from a good description
of the two lowest-lying $\pi_1$ states and identifying a good ``marker'', namely the axial vector channels, we 
transfer our assumptions to the other quarkonium systems and provide predictions for the masses of 
analogous states there. We see the lowest-lying $1^{-+}$ state in both charmonium and bottomonium
appear at about the same mass as the axial vector ground states. In summary, we have provided the hypothesis that,
if our model assumptions are sufficient to capture the necessary physics and if the two lowest-lying $\pi_1$ states
are well-represented by our setup, then one should expect to find analogous states also in the other quarkonium systems
as described in detail above. 

To broaden our set of results we will investigate the decay properties of such states in the near
future along the lines of previous studies of electromagnetic, leptonic, and hadronic transitions
in our approach, which have been proven feasible and successful phenomenologically, e.\,g., in 
\cite{Holl:2005vu,Bhagwat:2006pu,Mader:2011zf}.

\begin{acknowledgments}
We acknowledge helpful conversations with C.~Popovici, J.~Segovia, and R.~Williams.
This work was supported by the Austrian Science Fund (FWF) under project no.\ P25121-N27.
\end{acknowledgments}

\end{document}